\documentclass[aoas,nameyear,dvips]{arximspdf}
\usepackage{dcolumn}
\usepackage{graphicx}

\doi{10.1214/10-AOAS407}
\volume{5}
\issue{2A}
\pubyear{2011}
\firstpage{894}
\lastpage{923}

\makeatletter
\def\ul{\underline}

\newcolumntype{d}[1]{D{.}{.}{#1}}

\newcommand{\Prob}{\operatorname{Prob}}
\newcommand{\cT}{{\mathcal{T}}}
\newcommand{\bV}{\mathbf{V}}
\newcommand{\diag}{\operatorname{diag}}

\newcommand{\cov}{\operatorname{cov}}

\newcommand{\bY}{\mathbf{Y}}
\newcommand{\bt}{\mathbf{t}}
\newcommand{\bd}{\mathbf{d}}
\newcommand{\bB}{\mathbf{B}}
\newcommand{\bU}{\mathbf{U}}
\newcommand{\bE}{\mathbf{E}}
\newcommand{\gam}{\gamma}
\newcommand{\N}{{\mathcal{N}}}
\newcommand{\cP}{{\mathcal{P}}}
\newcommand{\cW}{{\mathcal{W}}}
\newcommand{\cJ}{{\mathcal{J}}}
\newcommand{\MGP}{{\mathcal{MGP}}}
\newcommand{\MN}{{\mathcal{MN}}}
\newcommand{\kron}{\otimes}
\newcommand{\vecv}{\operatorname{vec}}
\makeatother

\begin{document}
\begin{frontmatter}

\title{Automated analysis of quantitative image data using isomorphic functional mixed models,
with application to proteomics data}
\runtitle{ISO-FMM for image data}

\begin{aug}
\author[A]{\fnms{Jeffrey S.} \snm{Morris}\corref{}\thanksref{t1,t2,t3}%
\ead[label=e1]{jefmorris@mdanderson.org}\ead[label=u1,url]{http://works.bepress.com/jeffrey\_s\_morris}},
\author[A]{\fnms{Veerabhadran} \snm{Baladandayuthapani}\thanksref{t1}\ead[label=e2]{veera@mdanderson.org}},
\author[A]{\fnms{Richard C.} \snm{Herrick}\thanksref{t1}\ead[label=e3]{rherrick@mdanderson.org}},
\author[B]{\fnms{Pietro} \snm{Sanna}\thanksref{t2}\ead[label=e4]{psanna@scripps}}
\and
\author[A]{\fnms{Howard} \snm{Gutstein}\thanksref{t1,t2}\ead[label=e5]{hgutstein@mdanderson.org}}
\runauthor{J. S. Morris et al.}
\thankstext{t1}{Supported by National Cancer Institute (CA-107304).}
\thankstext{t2}{Supported by National Institute for Alcohol Abuse (AA-016157).}
\thankstext{t3}{Supported by Statistical and Applied Mathematical Sciences
Institute (SAMSI) Program on Analysis of Object Data (AOD).}
\affiliation{University of Texas MD Anderson
Cancer Center, University of Texas MD Anderson
Cancer Center, University of Texas MD Anderson
Cancer Center, Scripps Research Institute and University of Texas
MD Anderson~Cancer Center}
\address[A]{J. S. Morris\\
V. Baladandayuthapani\\
R. C. Herrick\\
H. Gutstein\\
University of Texas\\
\quad MD Anderson Cancer Center\\
Unit 1411, PO Box 301402\\
Houston, Texas 77230-1402\\
USA\\
\printead{e1}\\
\phantom{E-mail:\ }\printead*{e2}\\
\phantom{E-mail:\ }\printead*{e3}\\
\phantom{E-mail:\ }\printead*{e5}\\
\printead{u1}\\} %adresu isvedimo komanda gale!
\address[B]{P. Sanna\\
Scripps Research Institute\\
La Jolla, California\\
USA\\
\printead{e4}}
\end{aug}

% HISTORY:
\received{\smonth{12} \syear{2009}}
\revised{\smonth{5} \syear{2010}}

% ABSTRACT
%
\begin{abstract}
Image data are increasingly encountered and are of growing
importance in many areas of science. Much of these data are
\textit{quantitative image data}, which are characterized by intensities
that represent some measurement of interest in the scanned images.
The data typically consist of multiple images on the same domain and
the goal of the research is to combine the quantitative information
across images to make inference about populations or interventions.
In this paper we present a unified analysis framework for the
analysis of quantitative image data using a Bayesian functional
mixed model approach. This framework is flexible enough to handle
complex, irregular images with many local features, and can model
the simultaneous effects of multiple factors on the image
intensities and account for the correlation between images induced
by the design. We introduce a general \textit{isomorphic modeling
approach} to fitting the functional mixed model, of which the
wavelet-based functional mixed model is one special case. With suitable
modeling choices, this approach leads to efficient calculations and
can result in flexible modeling and adaptive smoothing of the
salient features in the data. The proposed method has the following
advantages: it can be run automatically, it produces inferential
plots indicating which regions of the image are associated with each
factor, it simultaneously considers the practical and statistical
significance of findings, and it controls the false discovery rate.
Although the method we present is general and can be applied to
quantitative image data from any application, in this paper we focus
on image-based proteomic data. We apply our method to an animal
study investigating the effects of cocaine addiction on the brain
proteome. Our image-based functional mixed model approach finds
results that are missed with conventional spot-based analysis
approaches. In particular, we find that the significant regions of
the image identified by the proposed method frequently correspond to
subregions of visible spots that may represent post-translational
modifications or co-migrating proteins that cannot be visually
resolved from adjacent, more abundant proteins on the gel image.
Thus, it is possible that this image-based approach may actually
improve the realized resolution of the gel, revealing differentially
expressed proteins that would not have even been detected as spots
by modern spot-based analyses.
\end{abstract}

% KEYWORDS
%

\end{frontmatter}

%s1 ###
\section{Introduction} \label{sec:intro}

Image data are increasingly encountered in
many areas of science and technology, including medicine, defense, robotics,
security, and materials science. Image analysis involves the extraction of
meaningful information from these data.

Some types of image analysis are performed subjectively by an expert
user who is trained to visually extract the important features from
the image. For example, a trained radiographer may inspect a CT scan
to determine whether a patient has a tumor, or a trained pathologist
may look at a scanned microscopic slide and determine the histology
of a tumor. Other types of image information can be automatically
extracted using a computer-based analysis of the digitized image
based on an expert systems approach. For example, face recognition
software can be used to identify an individual in an image, or
optical character recognition software can be used to ascertain
license plate numbers from still images taken at a toll booth. In
these examples, the data are digitized and pattern recognition is
used to perform discrimination, but the analysis is still
qualitative in nature because the information of interest is the
presence or absence of particular features in the image, not the
magnitudes of the pixel intensities themselves.

In other image data, the magnitudes of the digitized pixel
intensities actually represent an approximate quantification of some
measurement of interest. For example, in functional magnetic resonance
imaging (fMRI), magnetic images are
obtained for serial slices of the brain, and the pixel intensities
represent the amount of oxygenated blood flow to that part of the
brain, which is a surrogate measure for the brain activity level. In
2D gel electrophoresis (2-DE)-based proteomics, the proteomic
content of a biological sample is physically separated on a
two-dimensional polyacrimidic gel by its isoelectric point (pH) and
molecular mass. The gel is scanned to produce an image
characterized by spots that correspond to proteins present in the
sample. The intensities of the spots are rough measures of protein
abundance. We refer to this type of image data as \textit{quantitative
image data} (\textit{QID}), which is the primary focus of this paper.

A set of QID typically involves multiple scanned images from the
same individual and/or from different individuals, with intensities
observed over the same two-dimensional (or higher) domain. The
overall goal of this type of quantitative image analysis (QIA) is to
combine information across images to make statistical inferences
about populations or about the effects of certain interventions on
the populations represented in the images. One important specific goal
of QIA
is to identify which regions of the image differ significantly
across treatment groups or populations. For example, in fMRI we
might analyze images from individuals performing various functions
in order to determine which parts of the brain are typically active
during each activity, or we might wish to distinguish between
different populations of patients with respect to their brain
activity during a given activity, for example, in response to visual
stimulus for children with and without attention deficit disorder.
In proteomics, we aim to find which regions of the gel, and thus
which proteins, are differentially expressed between cases and
controls in a case-control study.

These image data sets are enormous in size and complex in nature,
presenting numerous challenges in terms of storing, managing,
analyzing and viewing the data. It is not uncommon to have
hundreds or thousands of images in a given data set, with each image
sampled on a grid of tens of thousands to millions of pixels.
Managing and viewing data sets of this size is difficult; performing
rigorous statistical analyses on the data is particularly
challenging. Before statistical analysis can be performed, the
raw, digitized images must undergo a number of processing steps,
including alignment, background correction, normalization,
denoising and artifact removal. These steps may involve
technology-specific methods, and must be done before any further
analysis takes place. We will not discuss preprocessing methods in
detail in this paper, but will assume the researchers have applied
suitable processing methods to the data before using the QIA methods
we describe.

\textit{Feature extraction vs. image-based modeling}: Researchers
frequently use a \textit{feature extraction} approach to analyze
QID. They are motivated by the premise that the relevant
information in the images is contained in well defined, discrete
features that can be extracted by computing numerical summaries
according to their estimated feature structure. The steps of a
feature extraction approach are to identify the salient features in
the images, quantify each feature for each individual, and then use
standard univariate and multivariate statistical methods to
determine which features are associated with the factors of
interest. For example, in fMRI, the preprocessed pixel intensities
can be integrated within predefined \textit{Regions of Interest
\textup{(}ROI\textup{)}}, for example, brain regions, and then these regions can be analyzed
to determine which regions are related to the underlying activity or
population. In 2-DE proteomics, a spot-detection algorithm
estimates distinct protein spots in the images. The protein spots
are quantified and then surveyed to determine which are
differentially expressed. This approach is computationally
efficient because it reduces the data from complex, high-dimensional
images to a vector of spot intensities, and can retain the relevant
information contained in the QID, provided that all the salient
features are properly detected and quantified.

The problem with this approach is that any information in the image
not contained in one of the feature summaries will be completely
lost to the analysis. In fMRI, there may be important differences
within subregions of the predefined regions of interest that could
be missed by integrating the entire region. In 2-DE, spot detection
methods are not perfect and may fail to detect some differentially
expressed proteins as distinct spots. One of the inherent dangers
of this problem is that the researchers may never be aware that they
missed anything---that there was information of significance in
their data that was missed because of inadequate feature extraction.

An alternative to feature extraction is to model the images in their
entirety using a suitable statistical modeling framework, which is a
challenging endeavor we call an \textit{image-based modeling}
approach. To be appropriate for image-based modeling, an analytic
method must possess the following three major characteristics: (1)
sufficient flexiblity and adaptability to accommodate the local
features that tend to characterize these complex, irregular data;
(2) the ability to appropriately borrow strength spatially across
the image; and (3) enough computational efficiency to be feasibly
applied to data sets of this magnitude. Some examples of recently
published image-based modeling methods include those of Reiss and
Ogden (\citeyear{ReiOgd2009}), who constructed a generalized linear model method for
image predictors, and Smith and Fahrmeir (\citeyear{SmiFah2007}), who analyzed fMRI
data using Bayesian variable selection on image pixels and an Ising
prior to model spatial correlation among the variable selection
parameters.
QID can be viewed as functional data with the two-dimensional domain
given by the rows and columns of the image and the range given by
the pixel intensities, and thus can be analyzed using a functional
data analysis [FDA, Ramsay and Silverman (\citeyear{RamSil1997})] approach.

Recent work on Functional Mixed Models (FMM) [Guo (\citeyear{Guo2002}), Morris et
al. (\citeyear{Morrisetal2003}), Morris and Carroll (\citeyear{MorCar2006}), Morris et al. (\citeyear{Morrisetal2006}), Morris et
al. (\citeyear{Morrisetal2008})] provides a general modeling framework useful for modeling
many types of functional data, but has not yet been adapted for use
with image data. In this paper we present a unified, Bayesian
image-based analysis approach for QID based on a version of the FMM
suitable for higher dimensional image data. The model fitting is
done using an \textit{isomorphic transformation} approach, which we
define and introduce in Section \ref{subsec:ISO-FMM}. The method
can simultaneously model the effects of multiple factors on the
images through fixed effects and can account for correlations
between images that are induced by the design through random effect
modeling. The isomorphic modeling approach results in efficient
calculations and, with suitable transformation, can accommodate
nonstationary features in the covariance matrices, and result in
adaptive smoothing and borrowing of strength across pixels in each
dimension while the inference is performed. The method yields
posterior probabilities of specified effect sizes that can be
interpreted as local false discovery rates [FDR, Benjamini and
Hochberg (\citeyear{BenHoc1995}), Storey (\citeyear{Storey2003})]. The posterior probabilities can be
used in Bayesian
inference to flag regions of the curves as significant while
considering both practical and statistical significance and
controlling the FDR. The software to implement this method can be
run automatically with little user input, and is efficient enough to
handle even very large image data sets. Although this method is
generally applicable to all QID, in this paper we focus on 2-DE
data. We show that this adaptive, image-based approach can find
results that would have been missed with standard spot-level
analyses, and may extract more protein information from the gels
than was previously known to be present.

In Section \ref{sec:app} we discuss image-based proteomics and
standard analysis approaches, and introduce the brain proteomics
data set that we consider in this paper. In Section
\ref{sec:methods} we describe the methodology: we
overview functional mixed models, describe our general isomorphic
approach to model fitting, and present the isomorphic functional
mixed model for higher dimensional image data. Also, we describe how
to conduct Bayesian FDR-based inference using the output data and
present an image compression
approach that can be used optionally to speed up calculations. In
Section \ref{sec:example} we apply this method to the brain
proteomics data set and compare and contrast results with a feature
extraction approach. We finish with a discussion of the implications
of our results for 2-DE proteomics and as a~general methodology for
quantitative image data analysis in Section \ref{sec:discussion}.

%s2 ###
\section{Image-based proteomics data} \label{sec:app}

%s2.1 ###
\subsection{Introduction to proteomics}

Over the past two decades,
advances in genomics have fueled increased interest in the field of
proteomics. Proteomics differs from genomics in that the former
field involves the direct measurement of proteins rather than their
precursors, genes and messenger RNA.
%The key advantage of proteomics is that
%proteins are the main effector molecules in cells, so may be more
%directly relevant to biological processes than gene expression values
%or genotypes.
%Their drawback is that they present more complex issues for analysis,
%with their 3D structure, large numbers of amino acids, and
%post-translational
%modifications.
%The are various types of proteomic studies have different goals that
%range from simply cataloging the proteins present in a sample, to
%studying the 3D structures of various proteins, to biomarker
%discovery.
Our focus is on the use of proteomics for biomarker
discovery, which involves the measurement of the relative abundance of
proteins across
different samples to determine which are differentially expressed
across groups or correlated to a factor
of interest. The proteins of interest can then be validated and further
studied for
possible clinical applications, for example, for early detection of cancer
or as markers of response to a particular cancer therapy. Various types
of proteomics data can be considered quantitative image data,
including liquid chromatography--mass spectrometry (LC--MS) and
2D gel electrophoresis (2-DE).

While LC--MS is growing in importance, the major workhorse
in biomarker discovery proteomics to date has
been 2-DE. The process of 2-DE involves
staining and denaturing the biological sample, running it through a
polyacrimidic gel, and separating the proteomic content of the
sample by isolectric point (pH) and then by molecular mass. The gel
is then digitally scanned to produce an image of the stained spots
that correspond to proteins present in the sample, which are double
indexed by their molecular mass and pH. The spots on the gel
physically contain the actual proteins, so protein identification is
easily accomplished by cutting out the spot, enzymatically digesting
it, and using MS--MS to ascertain its identity. One variant of 2-DE
that may yield more accurate relative quantifications is 2D
difference gel electrophoresis [DIGE, Lilley (\citeyear{Lilley2003}), Karp and Lilley
(\citeyear{KarLil2005})], which involves differentially labeling two samples with two
different dyes, loading them onto the same gel, and then scanning
the gel with two different lasers, each of which specifically picks
up one of the two dyes. When comparing two groups, paired samples
from each group can be run on the same gel, effectively conditioning
the gel effect out of the analysis. In more general problems, one
dye (the active channel) can be used for the primary sample and the
other dye (the reference channel) used on some common reference
material used on all gels so that it may serve as an internal
normalization factor.

2-DE has been criticized for various perceived limitations of the
technology, including its limited ability to measure proteins with
medium or low abundance or to resolve co-migrating proteins with
similar pH/mass combinations [Gygi et al.~(\citeyear{Gygietal200})]. Although the
technology itself may possess some technical limitations, a major
factor limiting the realized potential of 2-DE is a lack of
efficient and effective algorithms to process and analyze the gel
images. More effective analytic methods that better extract
proteomic information from the gel images may help the technology
more fully realize its potential.

%mainly for its ability to analyze low and medium abundance proteins.

%s2.2 ###
\subsection{Spot-based analysis of 2-DE proteomic data} %
Nearly all existing 2-DE gels are
analyzed using a feature extraction approach whereby spots are
detected and quantified for different gel images and then analyzed
to ascertain which are differentially expressed.
%Even though
%thousands of spots may be surveyed, proteomic analyses rarely
%adjusted for multiple testing until recently. Some researchers have
%emphasized the need for these adjustments, and have suggested
%approaches based on the false discovery rate (Gutstein and Morris
%2007, Karp, et al. 2007, Gutstein, et al. 2008).
The success of a feature extraction approach depends on
the effectiveness of the feature detection and quantification method
used and, until recently, the predominant approaches used for spot
detection and quantification had major problems. Traditional approaches
based on spot detection on individual gels followed by matching spots
across gels suffer from problems with missing data, spot detection
errors, spot matching errors and spot boundary estimation errors [Clark
and Gutstein (\citeyear{ClaGut2008}), Morris, Clark and Gutstein (\citeyear{MCG2008})]. The abundance
of these
errors may be partially responsible for some researchers concluding
the technology is ineffective. In recent years, alternative spot
detection strategies have been developed that mitigate these errors
to a degree, and include \textit{Pinnacle}, a method we have
developed [Morris, Clark and Gutstein (\citeyear{MCG2008}), Morris et al. (\citeyear{MCWG2010})],
and commercial packages
\textit{SameSpots} by Nonlinear Dynamics (Newcastle upon
Tyne, UK), \textit{Redfin Solo} by Ludesi (Malmo, Sweden) and
\textit{Delta2D} by Decodon (Greifswald, Germany). While
improving from past methods, these spot-based approaches are still far from
perfect, and, in particular, still have some difficulty resolving
distinct co-migrating proteins that are present in the same spot.
Thus, there may be more to gain by using an image-based modeling
approach.

\subsection{Image-based analysis approaches for 2-DE} Spot-based
approaches are almost
universally used for the analysis of 2-DE data. We know of only one
paper in the current literature that describes the application of an
image-based modeling approach. Faergestad et al. (\citeyear{Faergestadetal2007}) presented a
pixel-based method for pairwise analysis of 2-DE data that involves
the application of partial least squares regression (PLSR) to the
vectorized gel images (after preprocessing). Faergested et al. used
a jacknife procedure to conduct inference, repeatedly applying the
PLSR to each leave-one-out cross-validation sample and then
performing a $t$-test at each pixel using the cross-validation
regression coefficients as the data. In order to avoid flagging
pixels that were statistically but not practically significant, they
restricted their attention to pixels with a certain minimum standard
deviation across samples. The general image-based method we
introduce in this paper is not limited to pairwise inference; it can
account for correlation among images from the same subject or batch;
it performs adaptive smoothing as part of the estimation and
inference; and it yields rigorous unified FDR-based Bayesian
inference that simultaneously accounts for both statistical and
practical significance. Our method can be applied to any type of
image-based proteomics data, including LC--MS, 2-DE and DIGE.

%s2.4 ###
\subsection{Motivating example: Cocaine addiction brain proteomics
study} \label{subsec:intro_example}
The methods we develop in this paper are applied to proteomic
data from a neurobiology
study on cocaine addiction. The study aimed to identified
neurochemical changes in the brain that are associated with the
transition from nondependent drug use to addiction. The addiction
process is conceptualized as an increasing motivation to seek drugs,
resulting in increased drug intake, loss of control over drug
intake and compulsive drug taking. Previous studies suggest that
prolonged exposure to cocaine or opiate drugs leads to increased
self-administration and a~pronounced elevation in reward thresholds
[Leith and Barrett (\citeyear{LeiBar1976}), Kokkinidis, Zacharko and Predy (\citeyear{KZP1980}), Markou and Koob
(\citeyear{MaKoo1992}), Schulteis et al. (\citeyear{Schulteisetal1994})]. Neurochemical changes in parts of the
basal forebrain structure, the extended amygdala, parallel these
decreases in the function of the reward system [Parsons, Koob and Weiss~(\citeyear{PKW1995}),
Weiss et al.~(\citeyear{Weissetal1992}), Heinrichs et al. (\citeyear{Heinrichsetal1995}), Richter and Weiss (\citeyear{RichWei1999})].
These data suggest that substance dependence or addiction produces a~pronounced dysregulation of the brain's reward systems, and that
neurochemical changes in the extended amygdala may provide a
substrate for such dysfunction. The neurochemical changes may
involve cellular effects at the translational and post-translational
levels that alter protein expression and function, and thus may be
detected by proteomic analysis.

An animal study to investigate these concepts used a model developed
by Ahmed and Koob (\citeyear{AhmKoo1998}). The animal model was based on rats that
were trained to obtain cocaine by pressing a lever. Six rats were
given short durations of drug access (1 hour/day), and 7 rats were
given long durations of drug access (12~hours/day). The study
included 8 control rats. The rats were eventually euthanized, and
their brain tissue was harvested and microdissected to extract
various regions of the extended amygdala. Tissues from the brain
samples were then subjected to 2-DE to assess their proteomic
content. The goal was to compare and contrast protein expression
and modification associated with excessive levels of cocaine intake
and to compare tissues from animals given long versus short access
to the drug. The data we analyzed for this paper were obtained from
the the central nucleus region of the extended amygdala. The data
set contains a total of 53 gels from 21 rats, with roughly 2--3 gels
per rat.

%s3 ###
\section{Methods} \label{sec:methods}

In this section we review previous work on functional mixed models
and wavelet space modeling for 1D functional data, and then discuss
how this approach can be used with other transformations,
that is, not just wavelets. Thereafter, we describe how to adapt this
method to model image data and discuss how to perform rigorous
FDR-based Bayesian inference from its output.

%s3.1 ###
\subsection{Functional mixed models and wavelet-based modeling} \label{subsec:FMM}

For back-\break ground, here we describe the wavelet-based functional mixed
models method (WFMM) of
Morris and Carroll (\citeyear{MorCar2006}). Suppose we observe a sample of $N$ curves
$Y_i(t), i=1,\ldots,N$, each defined
on a compact set $\cT$. The FMM is given by
\begin{equation}\label{eq:FMM1}
\bY(t)=X \bB(t) + Z \bU(t) + \bE(t) ,
\end{equation}
where $\bY(t)=\{Y_1(t), \ldots, Y_N(t)\}'$ is a vector of observed
functions, ``stacked'' as rows. Here, $\bB(t) = \{B_1(t), \ldots,
B_p(t)\}'$ is a vector of fixed effect functions with corresponding $N
\times p$ design matrix $X$, $\bU(t) = \{U_1(t), \ldots, U_m(t)\}'$ is
a~vector of random effect functions with corresponding $N \times m$
design matrix $Z$, and $\bE(t) = \{E_1(t), \ldots, E_N(t)\}'$ is a
vector of functions representing the residual error processes. The
effect functions measure the partial effect of the corresponding
covariate at position $t$ of the functions. The set of random effect
functions $\bU(t)$ is a realization from a
(mean zero) multivariate Gaussian process with $m \times m$
between-function covariance matrix $P$ and within-function
covariance surface $Q(t_1,t_2)$, denoted by $\bU(t) \sim\MGP(P,Q)$
and implying that $\cov\{U_b(t_1),U_{b'}(t_2)\}=P_{bb'} Q(t_1,t_2)$.
%We constrain $P$ to be a correlation matrix, which is sufficient (but
%not
%necessary) to ensure identifiability.
The residual errors are assumed to follow $\bE(t) \sim\MGP(R,S)$,
independent of $\bU(t)$.
The random effect or residual error
portions of the model can be stratified to allow covariances indexed
by some factor, $Z \bU(t)=\sum_{h=1}^H Z_h \bU_h(t)$ with $\bU_h(t)
\sim\MGP(P_h,Q_h)$ or $\bE(t) = \sum_{c=1}^C \bV_c \bE_c(t)$ with
$\bV_c$ a vector whose $i$th element is 1 if curve $i$ is from
stratum $c$ and $\bE_c \sim\MGP(R_c, S_c)$.
%A commonly used
%special case involves independent and identically distributes (iid)
%random effect functions with $P=R=I$.
%This model has the following characteristics: good flexibility, the
%ability to simultaneously model functional effects for multiple
%covariates of interest, and, through the random effect structure,
%the ability to account for various covariance structures between and
%within functions induced by the experimental design.

In practice, observed functional data are sampled on some
discrete grid. Assuming all observed functions are sampled on the same
fine grid $\bt= (t_1, \ldots, t_T)$, the discrete version of (\ref{eq:FMM1})
is
\begin{equation}\label{eq:FMM2}
Y = X B + Z U + E,
\end{equation}
where $Y$ is an $N \times T$ matrix of observed curves on the grid
$\bt$, $B$ is a $p \times T$ matrix of fixed effects, $U$ is an $m
\times T$ matrix of random effects, and $E$ is an $N \times T$ matrix
of residual errors.
%As defined above, $X$ is an $N \times p$ matrix
%and $Z$ is an $N \times m$ matrix, and the two are the design matrices
%for the fixed and random effect functions, respectively.
Following Dawid (\citeyear{David1981}), $U$ follows a matrix normal distribution
with $m \times m$ between-row covariance matrix $P$ and $T \times T$
between-column covariance matrix $Q$, which we denote by $U \sim
\MN(P,Q)$, implying $\cov(U_{ij},U_{i'j'})=P_{ii'}Q_{jj'}$.
The residual error matrix $E$ is assumed to be $\MN(R,S)$.
The within-random effect curve covariance surface $Q$ and
residual error covariance surface $S$ are $T \times T$
covariance matrices that are discrete approximations of the
corresponding covariance surfaces in $\cT\times\cT$.

Morris and Carroll (\citeyear{MorCar2006}) used a wavelet basis modeling approach to fit
the model
(\ref{eq:FMM2}), which involves three steps. First, a fast
algorithm called the discrete wavelet transform [$\operatorname{DWT}$, Mallat (\citeyear{Mallat1989})]
is applied to each of the~$N$ observed functions on grid $\bt$ to
yield a vector of $T$ wavelet coefficients for each function, effectively
rotating the data axes to transform the data into the wavelet space.
Second, a Markov chain Monte Carlo (MCMC) procedure is used to
obtain posterior samples from a wavelet-space version of model~%
(\ref{eq:FMM2}). The wavelet-space covariance matrices for the
random effect functions and residual errors are modeled as diagonal,
but with different variances for each wavelet coefficient, and spike-slab
priors are assumed on the fixed effects' wavelet coefficients. These
assumptions are parsimonious, yet accommodate nonstationary features in
the data-space covariance matrices $Q$ and $S$ and induce
adaptive regularization of the fixed and random effect functions,
$B_a(t)$ and $U_b(t)$
[Morris and Carroll (\citeyear{MorCar2006})]. Third, the inverse $\operatorname{DWT}$ is applied to the
the posterior samples of the wavelet-space
parameters to yield posterior samples of the parameters in the data-space
model (\ref{eq:FMM2}), which can be used to perform Bayesian inference.
\subsection{Isomorphic modeling of functional mixed models (ISO-FMM)}
\label{subsec:ISO-FMM}

The WFMM is just one example of a general
approach to fitting functional mixed models we call an
\textit{isomorphic} approach (ISO-FMM), which we introduce here. The same
basic three-step approach underlying the WFMM can be applied using
isomorphic transformations not involving wavelets if desired. We define
an \textit{isomorphic transformation} as one that preserves all of the
information in the original data, that is, is invertible. More
precisely, given row vector $\bf y\in\Re(T)$, we say a
transform $f\dvtx\Re(T)\rightarrow\Re(T)$ is
\textit{isomorphic} if there exists a reverse transform $f^{-1}$
such that $f^{-1}\{f(\bf y)\}=\mbox{\bf y}$. The wavelet
transform is
isomorphic because $\operatorname{IDWT}(\operatorname{DWT}(\bf y))=\bf y$, but isomorphic
transformations can be constructed in other ways as well, for
example, by using other basis functions including Fourier bases,
spline bases and certain empirically determined basis functions
like functional principal components.

%and describe the scope of transformations that could be used,
%discussing necessary and desirable properties.
Suppose we observe $N$ functions all on the same fine grid $\bt$ of
length $T$, resulting in $N \times T$ data matrix $Y$ whose rows are
the observed functions and columns index the grid locations. The
following steps describe the general steps of a Bayesian
implementation of the ISO-FMM for functional data:
\begin{enumerate}
\item Transform each of the rows of $Y$ using an isomorphic
transformation $f$, represented as $D=f(Y)$, with $f(\cdot)$ applied
to a matrix implying here the transform $f$ is applied separately
to each row.
%%%
%%% Problem: what is the continuous analog to this?
%%%
Rather than indexing positions within the curve, the columns of $D$
will index items in the transformed space, for example, basis coefficients.
We can think of the induced functional mixed model in the
transformed space
%(\ref{eq:FMM3}),
with the columns of $D, B^*=f(B),
U^*=f(U)$, and $E^*=f(E)$ indexing coefficients in the alternative
space. We refer to this as the \textit{transformed-space FMM}.

\item Apply an MCMC procedure to the transformed-space FMM to
obtain posterior samples of all of its parameters. This requires
specification of (a)~parsimonious assumptions on the covariance
matrices $Q^*$ and $S^*$ that are sufficiently flexible to capture
important features of $Q$ and $S$, and (b)~a~prior distribution on
the fixed effects in the transformed-space FMM to induce effective
regularization of the fixed effect functions $B^*$.

\item Apply the inverse isomorphic transform $f^{-1}$ to the posterior
samples of the
functional quantities in the transformed-space FMM to obtain posterior
samples from
the original data-space FMM (\ref{eq:FMM2}), where Bayesian inference is
performed.
\end{enumerate}
This approach could also be applied in a frequentist context. That
would involve fitting the transformed-space model with some explicit
roughness penalties in appropriate places, for example, the fixed and
random effect functions, to induce adaptive smoothing, and then
transforming the estimated quantities back to the data space. This
would easily yield estimates, but more work would need to be done
to obtain inferential quantities.

The use of an isomorphic transformation ensures that the
representation in the transformed data retains all of the
information contained in the original data, that is, is ``lossless,'' and
thus any basis coefficients can be considered as transformed raw
data rather than estimated parameters. Thus, the transformed-space
model is isomorphic to the data-space model. This allows us to
perform the modeling in the transformed space, where it may be possible
to perform modeling and regularization more parsimoniously and
conveniently, and yet obtain valid inference in the data-space model,
where the
parameters are more clearly interpretable.

For many isomorphic transformations, it is possible to assume
parsimonious structures for $Q^*$ and $S^*$ in the basis space and
still accommodate a~rich class of structures for data space within
curve covariance matrices $Q$ and~$S$. For example, using a Fourier
transform, any stationary covariance matrix can be represented by
uncorrelated Fourier coefficients, so diagonal~$Q^*$ and $S^*$ are
fully justified if we are willing to assume stationarity in~$Q$ and~$S$.
Diagonal assumptions on $Q^*$ and~$S^*$ allow the
transformed-space FMM to be fit one column at a time, making the
procedure highly parallelizable and reducing the memory requirements
of the software. This assumption may also be justifiable in some
empirically-determined basis spaces such as FPC. For wavelets, the
whitening property of the transform makes diagonal $Q^*$ and~$S^*$ a
reasonable working assumption that accommodates many commonly
encountered nonstationary features. For a given isomorphic
transformation, one must decide what parsimonious assumptions are
reasonable in the basis space, and carefully consider what
constraints these assumptions induce in the data space.

Another advantage of transformed-space modeling is that for many
isomorphic transforms, there are natural prior distributions on
basis space coefficients that can induce regularization of the
functional effects in the model and effectively act like roughness
penalties. For example, with wavelets, a~sparsity prior that has
a spike at zero and medium-to-heavy tails like a spike-slab prior
%(\ref{eq:spikeslab}) induces nonlinear shrinkage toward zero,
leads to adaptive regularization of the underlying
effect function. Spline bases are frequently regularized by second-order penalties,
which can be induced by a Gaussian prior. First-order penalties can be induced by double-exponential priors.

Although orthonormal linear isomorphic transformations are
convenient to use because they represent a simple rotation of the
axes, they are not the only possibility. The transform does not have to
be orthonormal or
even linear. With an orthonormal transform, i.i.d. white noise has the
same distribution and total energy in both the data
and basis space, but these are not necessary properties for the FMM.
With linear transforms [$f(Y)=YW'$ for some matrix $W'$], a Gaussian
model in the data space induces a~Gaussian model in the transformed
space, and vice-versa, but this is also not absolutely necessary for
valid modeling. For example, one could specify a Gaussian model in
the transformed space that is used for the fitting, and this
would correspond to some non-Gaussian model in the data space that
might not have a simple closed form, but which could still be a valid and
reasonable data-space likelihood.

If the set of functions jointly have a very sparse representation
in the chosen basis space, it may be possible and advantageous to
use an approximately isomorphic transformation of lower dimension
that still retains almost all of the information for the original
functions. We describe a way to perform this \textit{compression} in
the multiple function context in Section \ref{sec:compression}.

Many methods in the existing statistical literature use a basis
function approach to represent functions or vectors, but rather than
transformed data the coefficients are typically treated as parameters
to estimate and the transforms are not isomorphic but lower rank
projections. There are some methods in the current statistical literature
that effectively use an isomorphic modeling approach [e.g., wavelet
regression, Clyde, Parmigiani and Vidakovic (\citeyear{CPV1998}); spectral analysis of stationary
time series, Diggle and Al Wasel (\citeyear{DigWas1997}); nonisotropic modeling of
geostatistical data, Sampson and Guttorp (\citeyear{SamGut1992})]; however, to our knowledge,
this has not been discussed previously as a general modeling strategy.
Our intended contributions here are to (1) explicitly offer an isomorphic
approach as a general modeling strategy and (2) apply this approach
to functional mixed modeling.

%s3.3 ###
\subsection{ISO-FMM for quantitative image data} \label{subsec:image_FMM}

In this section we introduce a~functional mixed model for image data,
describe how to model image data using our isomorphic transformed-space
approach, and provide
implementation details using higher dimensional wavelet transforms.
Even though these results
hold generally for higher dimensional images, we present the results
for 2D images for ease of exposition.

%s3.3.1 ###
\subsubsection{Functional mixed models for quantitative image data}
Suppose we have a
sample of $N$ images, $Y_i, i=1, \ldots, N$, with each $Y_i$ a
$T_1 \times T_2$ matrix containing the image intensities sampled on a
regular, equally-spaced two-dimensional grid
$(\bt_1, \bt_2)$ with $\bt_1=(t_{11}, \ldots, t_{1T_1})'$ and $\bt
_1=(t_{21}, \ldots, t_{2T_2})'$.
A~functional mixed model for these image data, with $(t_1, t_2)$ a coordinate
on the grid, can be written as
\begin{equation}\label{eq:Image_FMM}
Y_i(t_1,t_2) = \sum_{a=1}^p X_{ia} B_{a}(t_1,t_2) + \sum_{b=1}^m
Z_{ib} U_{b}(t_1,t_2) + E_i(t_1,t_2),
\end{equation}
where $B_a$ and $U_b$ are fixed and random effect images,
respectively, which measure the effects of scalar fixed or random effect
covariates on the corresponding location of the image $Y$, and $E_i$
contains the residual error images. The $U_b$ and $E_i$ are mean zero Gaussian
processes defined on the surface, with corresponding between-image covariance
matrices $P$ and $R$, respectively, and four-dimensional within-image
covariance surfaces $Q(t_1,t_2,t'_1,t'_2)$ and $S(t_1,t_2,t'_1,t'_2)$
summarizes the covariance between locations $(t_1,t_2)$ $(t'_1,t'_2)$
of the random effect and residual error images, respectively.

Let each image be represented by a row
vector of length $T_1*T_2$, $\mathbf{y}_i=\{\vecv(Y_i)\}'$, where
$\vecv$
is the column stacking vectorizing operator.
% with $\vecv(A)=[A_1', A_2', \ldots, A_{T_2}']'$ for the $T_1 \times
%T_2$ matrix $A=[A_1
%A_2, \ldots A_{T_2}]$.
If we let~$Y$ be the $N \times T(=T_1*T_2)$ matrix whose rows contain
the vectorized images, then the discrete image mixed model can be
written as
\begin{equation}\label{eq:IFMM}
Y^I = X B^I + Z U^I + E^I,
\end{equation}
with each row of $B^I$ and $U^I$ containing one of the vectorized fixed
or random effect images, respectively, that measure the effect of a
scalar fixed or random effect covariate on the corresponding
location of the image, and with the rows of $E^I$ containing the vectorized
``residual error images.'' The columns index the pixels in the image.
The superscript ``$I$'' simply is a~reminder that
these quantities are based on images. As before, we assume that $U^I
\sim
\MN(P,Q)$ and $E^I \sim\MN(R,S)$, with $P$ and $R$ being $m \times m$
and $N \times N$ matrices defining covariances between images, and
$Q$ and $S$ being $T \times T$ within-function two-dimensional
covariance matrices for the random effects and
residuals that model the covariance between different positions within the
images. For\vspace*{-1pt} example, $Q\{t_1+(t_2-1)*T_1, t_1^\dagger+ (t_2-1)^\dagger
*T_1\}$ describes the
covariance between $U^I_b(t_1,t_2)$ and $U^I_b(t_1^\dagger,
t_2^\dagger)$. Note that any reasonable structure on these
within-image covariance matrices should not just model the
autocovariance based
on the proximity within the vector $\mathbf{y}_i$, but rather the
proximity within the higher
dimensional image $Y_i^I$, that is, in all dimensions.

%s3.3.2 ###
\subsubsection{ISO-FMM for quantitative image data} The ISO-FMM
approach for 1D functions
described in Section \ref{subsec:ISO-FMM} can be applied to QID, as well,
using isomorphic transforms and inverse transforms that operate on the
higher dimensional functions, for example, images. The covariance
assumptions in the transformed space
and the regularization prior distributions should be chosen to induce
appropriate spatial correlation, adaptive smoothing and borrowing of strength
in all dimensions.

The isomorphic transform $f$ in the image space will map the $T=T_1
\times T_2$ pixels to
a set of alternative transformed-space coefficients, $D^I_i=f(Y^I_i)$.
This transform can be constructed
a number of different ways. One natural way is to take tensor products of
suitable 1D transforms, leading to a~\textit{separable} transform.
It is also possible to use special bases constructed for
image data. Depending on the resolution of the images and transform
used, computational
feasibility can become an issue because the transform will have to be applied
to all $N$ observed images as well as to all $p$ fixed effect images
for each
of $M$ posterior samples. After transformation, a model is then
proposed in the transformed
space.

If the transform is linear, then the Gaussian assumptions from (\ref{eq:IFMM}) hold in both the
data and transformed space, and our transformed-space model is given by
\begin{equation}\label{eq:IFMM2}
D^I = X B^{I*} + Z U^{I*} + E^{I*},
\end{equation}
with the rows of $D^I, B^{I*}=f(B^I), U^{I*}=f(U^I),$ and
$E^{I*}=f(E^I)$ containing the transformed representations
for each of the corresponding image-based quantities in (\ref{eq:IFMM}), with $U^{I*} \sim MN(P,Q^*)$
and $E^{I*} \sim MN(R,S^*)$. As before, $f(A)$ for some matrix $A$ means
applying the transformation $f$ sequentially
on the rows of~$A$. If a separable linear transform is used, then the linear
transform matrix for the vectorized images can be explicitly defined as
follows. Suppose
we obtain a matrix of coefficients $D^I_i$ from the sampled~ima\-ge~%
$Y^I_i$ by applying
a linear transform~$W_1$ to the rows of the image and $W_2$ to the
columns, that is,
$D^I_i=W_1 Y^I_i W_2'$. This transformation can be explicitly
represented as
$\bd_i = \mathbf{y}_i \cW'$, where $\mathbf{y}_i=\vecv
(Y^I_i)'$ and $\bd_i = \vecv(D^I_i)$
are the vectorized image and coefficient matrix, respectively, $\cW
'=(W_2 \kron W_1)$ is the linear
transformation matrix, and $\kron$ is the Kronecker product.
This representation makes it easy to explicitly see the connections between
the data-space and transformed-space matrix models (\ref{eq:IFMM}) and
(\ref{eq:IFMM2}) for the QID context
as $D^I=Y^I\cW'$, $B^{I*}=B^I\cW'$, $U^{I*}=U^I\cW'$, and
$E^{I*}=E^I\cW'$, and $Q^*=\cW' Q \cW$ and $S^*= \cW' S \cW$.
If $W_1$ and $W_2$ are orthogonal, then it follows that $\cW$ is also
orthogonal.
%Note that this structure can account for spatial correlations in both
%dimensions.
These results generalize to general $r$-dimensional functions stacked
as vectors using $\cW'=(W_r \kron W_{r-1}
\kron\cdots\kron W_1)$.

%s3.3.3 ###
\subsubsection{Implementation details using wavelets} The same
properties that make wavelet bases
convenient for isomorphic modeling in 1D functional~da\-ta (fast
calculations, compact support,
whitening property, joint frequency--time representation, sparse
representations for broad classes
of data) also~ma\-ke them useful for modeling QID. Here, we will describe the
implementation details for ISO-FMM using higher dimensional wavelet
transforms to construct the
isomorphic transformations, which involves three factors: choice of
transform, specification of
covariance structure, and regularization prior.

There are various ways to construct isomorphic transforms for image
data using wavelet bases. These
transforms can be separable or nonseparable. A~separable, or \textit{rectangular}, transform is easily constructed
by applying the 1D $\operatorname{DWT}$ separately to each row and each column of the
image. As mentioned in Section \ref{subsec:FMM},
after applying the wavelet transform to a vector of data, the resulting
wavelet coefficients are
double-indexed by scale $j=1, \ldots, J$ and location $k=1, \ldots,
K_j$. If we apply a separable 2D wavelet transform, each
coefficient is quad-indexed by row scale~$j_1$ and location $k_1$, and
column scale~$j_2$ and location
$k_2$.

Nonseparable transforms can also be used. Although they are not
represented as simple tensor products of
1D transforms, they are constructed using linear operators, and so
still represent a linear transformation.
The most commonly used nonseparable wavelet transform is a \textit{square} transform. This type of
decomposition yields three types of wavelet coefficients at each scale
$j=1, \ldots, J$, corresponding to horizontal,
vertical and diagonally-oriented wavelet bases. In this case, the
wavelet coefficients are triple-indexed by
scale ($j=1, \ldots, J$), orientation \{$l=1$ (row details), 2 (column
details), 3~(2D details)\}
and location ($k=1, \ldots, K_{jl}$). The square wavelet transform
tends to better model local behavior and
leads to more parsimonious representations than the rectangular
transform, and so is commonly used in practice.
With the basis functions aligned with the principal axes (horizontal,
vertical and diagonal), a disadvantage
of the square transform is that sometimes
it does not efficiently represent smoother contours or features of the
images that do not align
with the principal axes [Do and Vetterli (\citeyear{DoVet2001})]. This leads to less
effective adaptive smoothing for
images with these types of features. Other 2D wavelet transforms have
been constructed for this purpose
and could be used in place of the square transform, for example,
curvelets [Candes and Donoho (\citeyear{CanDon2000})], contourlets [Do and Vetterli (\citeyear{DoVet2005})]
or qincunx wavelets [Feilner, Van De Ville and Unser (\citeyear{FVU2005})]. We choose to use the
square nonseparable wavelet transform for
2-DE data because the key features of the images, the spots, are
aligned with the horizontal and vertical axes and
so should be well represented by them. We found them to be more
efficient than the rectangular separable transform which
contains many long, thin basis functions constructed by combining a low
frequency basis in one dimension (small $j$) and
high frequency basis in the other (large $j$).

Again, motivated by the whitening property of the wavelet transform, we
model the wavelet coefficients as
independent, that is, $Q^*=\diag(q_{jlk})$ and $S^*=\diag(s_{jlk})$,
allowing each coefficient triple-indexed
by its scale $j$, orientation $l$ and location $k$ to have its own
variance component. The independence leads to
parsimonious modeling, while the heteroscedasticity accommodates
nonstationary spatial features in the data space
matrices $Q$ and $S$. In Supplementary Material, we illustrate through
plots  and movies [Morris (\citeyear{Morris2010})]
the effective spatial covariance structures of $Q$ and $S$ induced by
independent heteroscedastic wavelet space models
for our 2-DE data. It accommodates spatial
covariance in all directions, based on proximity horizontally,
vertically and diagonally, and the strength of this
spatial covariance is allowed to vary across different parts of the
image. This adaptive handling of spatial correlation
is important in 2-DE, since we expect strong autocorrelation within
spots that rapidly falls off outside of the spot, and we
expect a~more slowly decaying autocorrelation in nonspot background
regions of the gel. Further, the structure allows different
image-to-image variances for different pixels in the image, which is
important to obtain accurate pixelwise
inference, since we expect different protein spots to have different variances.
%including different variances and different degrees of autocorrelation
%across the image and in different directions.
%This structure models autocorrelation in all directions, based on
%proximity horizontally, vertically, and diagonally,
%and
These principles generalize to higher dimensional images when the
corresponding higher dimensional $\operatorname{DWT}$ is used for transformation.

We assume the spike-Gaussian slab prior on the wavelet coefficients for the
fixed effects, which is written as
\begin{eqnarray}\label{eq:spikeslab2}
B^{*}_{ajlk} &=& \gam^{*}_{ajlk} \N(0,\tau_{ajl}) + (1-\gam
^{*}_{ajlk}) I_0,\nonumber\\ [-8pt]\\ [-8pt]
\gam^{*}_{ajlk} &=& \operatorname{Bernoulli}(\pi_{ajl}), \nonumber
\end{eqnarray}
with regularization parameters $\pi$ and $\tau$ indexed by covariate
$a$, scale $j$ and orientation $l$, and
estimated from the data using the empirical Bayes procedure similar to
Morris and Carroll (\citeyear{MorCar2006}),
as detailed in a supplementary article [Morris (\citeyear{Morris2010})]. This induces
adaptive smoothing of the fixed effect images $B_a(t_1,t_2)$.
By indexing the parameters by covariate, we allow for different
regularization parameters for different fixed effect
images, and by indexing by scale $j$ and orientation $l$, we are able
to naturally accommodate different degrees of smoothness
horizontally, vertically and diagonally within the fixed effect images.

After specifying vague proper priors on the variance components, we are
left with a fully specified Bayesian model
for the transformed-space FMM~(\ref{eq:FMM2}). We use a Markov chain
Monte Carlo (MCMC) procedure
to obtain posterior samples of the transformed-space fixed effect
functions $B^*$,
and then apply the inverse 2D wavelet transform to them to obtain
posterior samples
of the data-space fixed effect functions $B$, which are used for
Bayesian inference.
The MCMC details are presented in the supplementary article by Morris (\citeyear{Morris2010}).

%s3.4 ###
\subsection{Bayesian FDR-based inference} \label{subsec:FDR}
Given the posterior samples of\break $B_{a}(t_1, t_2)$, the fixed effect
image describing the effect of covariate $X_a$ on
the images as a~function of position $(t_1, t_2)$, we can perform
Bayesian inference to flag significant regions of
the curves by extending the approach used in Morris et al. (\citeyear{Morrisetal2008}), as follows.

First, we must define the effect size that is of practical
significance, say,~$\delta$. For example, if the image
intensities are modeled on a $\log_2$ scale, then $\delta=1$ would
correspond to a two-fold difference. From the posterior
samples of $B$, we can compute the posterior probability of an effect
size of at least~$\delta$, $p_a^\delta(t_1, t_2)=\Prob\{|B_a(t_1,
t_2)|>\delta\}$, which
can be plotted in what we call a~\textit{probability discovery image},
and define significant regions of the image as those with
$p^\delta_a(t_1, t_2)>\phi$ for some threshold $\phi$. The
quantities $1-p^\delta_a(t_1, t_2)$ can be considered $q$-values, or
estimates of the
local false discovery rate [Storey (\citeyear{Storey2003})], as they measure the
probability of a false positive if position $(t_1, t_2)$ is called a
``discovery,'' defined as a region in the image with at least $\delta$
effect size.

The significance threshold $\phi$ can be determined using classical
Bayesian utility considerations such as those of
Mueller et al. (\citeyear{MPRR2004}) based on the elicited relative costs of false
positive and false negative errors.
Alternatively, it can be set to control the average Bayesian FDR, in
the same manner as in Morris et al. (\citeyear{Morrisetal2008}).
For example, suppose we are interested in finding the threshold value
$\phi_\alpha^\delta$ that controls the overall average
FDR at some level $\alpha$ of the original image on a continuous
domain in the Lebesgue sense,
meaning we expect the ratio of Lebesgue measures of the falsely
discovered regions to regions flagged as discoveries to be no more than
$\alpha$.
When our interest is on the discrete grid of pixels sampled in the
observed image, we can estimate this threshold as follows. We drop the
index $a$ from all quantities to declutter the notation.
For all image pixel locations $(t_{1j}, t_{2j}), j=1, \ldots, T$, in the
vectorized probability of discovery image
$\ul{p}^\delta=[p^\delta_{j}; j=1, \ldots, T]= \vecv\{
p^\delta(t_1, t_2)\}, $
we first sort $p^\delta_{j}$ in
descending\vspace*{-2pt} order to yield $p^\delta_{(j)}, j=1, \ldots, T$.
Then $\phi^\delta_\alpha=p^\delta_{(\xi)}$, where $\xi=\max\{
j^*\dvtx j^{*-1}\sum_{j=1}^{j^*} \{1-p^\delta_{(j)}\} \le
\alpha\}$, which is the maximum index for which\vspace*{-1pt} the cumulative average
of the sorted local false discovery rates $(1-p^\delta)$
is less than or equal to $\alpha$.
The set of image regions $\cT^\delta_\alpha=\{(t_1, t_2)\dvtx p^\delta
(t_1,t_2)>\phi_\alpha^\delta\}$ are then flagged as ``significant,''
based on an effect size of $\delta$ and an average Bayesian FDR of
$\alpha$.
In 2-DE, a map of these image regions can be forwarded to the
spot-cutting robot in order to
cut out these regions of the gel for protein identification.

%s3.5 ###
\subsection{Image compression to speed computations} \label{sec:compression}

The ISO-FMM approach described in Section \ref{sec:methods} involves
transforming the observed functions or images into the transformed space
and modeling \textit{all} items in the transformed space, for example,
all basis coefficients.
Our approach and software are sufficiently computationally efficient
enough to
perform this procedure, even for quite large data sets. However, if the
chosen transformation leads to sparse representations of the observed images,
it may be possible to use a virtually ``lossless'' approach modeling a
subset of the
coefficients and to save a great deal of computational time and memory
overhead. If the transformation leads to a sparse representation,
then most of the basis coefficients are near zero for all images
and they could be left out of the modeling with very little practical
effect on the final results, effectively compressing the observed
images and all images in the FMM. For example, wavelets lead to
sparse representations of many classes of functional and image data
and are routinely used in signal compression applications, including
JPEG images and MPEG video.

In this section we introduce a compression method that selects which
coefficients to
include in the model in order to preserve a minimum percentage of the
total energy for
all images in the data set. This method can be used to plot the minimum
total energy vs.
number of coefficients, to help the user mitigate the trade-off between
information and
compression. This approach could also be used with criteria other than
total energy.

After transforming each of $N$ vectorized images to the transformed
space with~$T$ coefficients, we are left with the $N \times T$ matrix $D^I$ whose
rows $i=1, \ldots, N$ correspond to the images and columns $j=1,
\ldots, T$
correspond to the basis coefficients. For each row of $D^I$, we square
the coefficients,
sort them in decreasing order, and then compute the
relative cusum $C_{ij}$ for each coefficient $j$. The quantity $C_{ij}$
represents the proportion of
total energy preserved for curve~$i$ if only coefficients of magnitude
$|D^I_{ij}|$
and larger are retained. Define the set $\cJ_\cP=\{j\dvtx C_{ij}>\cP\mbox
{ for all } i=1, \ldots, N$\}
of size $T^*_\cP=\Vert\cJ_\cP\Vert$ to contain the indices of the minimal
set of coefficients
that must be kept to preserve $100 \cP\%$ of the total energy for each
image, with
complementary set $\cJ_\cP'$ containing the remaining coefficient indices.
In the transformed-space FMM, only the $T^*_\cP$ coefficients $j \in
\cJ_\cP$
would actually be modeled, and zeros would be substituted for the
regions of
$B^*, U^*, E^*, Q^*$ and $S^*$ corresponding to $j \in\cJ_\cP'$. One
can vary $\cP$ and
plot $\cP$ vs. $\Vert\cJ_\cP\Vert$ in what we call a \textit{compression
plot}---a useful tool
in deciding how much compression to do. This plot is a multiple-sample
analog to the scree
plot, a~commonly used tool in principal components analysis. Depending
on the image features
and transformation used, extremely high compression levels (100:1 or
greater) can
retain virtually all information contained in the raw images.
Note that these compression ratios also approximate the savings
in memory overhead required to run the MCMC procedure. Thus, this near
isomorphic
approach may be preferable to the full isomorphic approach modeling all
coefficients.

%s4 ###
\section{Application to brain proteomics data} \label{sec:example}

In this section we apply the wavelet-based ISO-FMM for quantitative
image data
described in Section \ref{sec:methods} to the brain proteomics data
set introduced in
Section \ref{subsec:intro_example}.

%s4.1 ###
\subsection{Methods} \label{subsec:example_methods}

\textit{Gel image preprocessing.} As described in Section \ref
{sec:methods}, we obtained
a total of 53 2D gel images from a total of 21 rats. We used one gel as
a~reference and
registered the other 52 gels to that reference in order to get the
protein spots
aligned across images using RAIN [Dowsey, Dunn and Yang (\citeyear{DDY2008})]. Then, we
cropped each registered
gel image within the same $646 \times861$ region to exclude parts of
the gel that were
either corrupted or did not appear to contain any proteins.
From each image we estimated and removed a spatially-varying local background
by subtracting from each pixel intensity the minimum value within a
square formed
by a window of $+\mbox{/}- 100$ around that pixel in horizontal and vertical
directions. We
normalized the image by dividing by the total sum of all
background-corrected pixel
intensities on the gel. We conducted both steps as described in Morris,
Clark and
Gutstein (\citeyear{MCG2008}). The resulting normalized intensities were then $\log
_2$ transformed
to yield the images $Y_i$ used for the downstream quantitative analyses.

\begin{figure}

\includegraphics{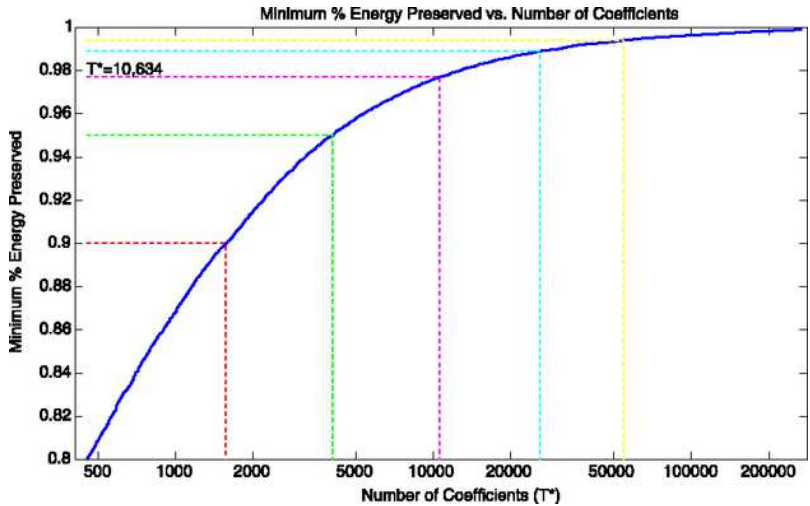}
\vspace*{-5pt}
  \caption{Compression plot: Plot of minimum proportion
  of energy preserved for EACH image vs. number of wavelet
  coefficients ($T^*$) for example 2-DE data set.}
  \label{fig:compression_plot}
  \vspace*{-6pt}
\end{figure}

\begin{figure}

\includegraphics{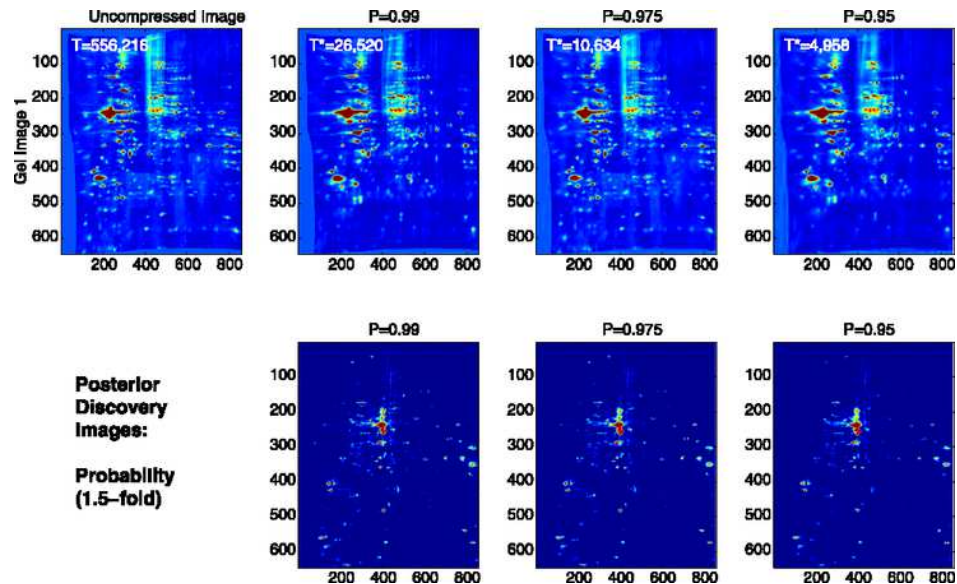}

  \caption{Illustration of compression: Heatmap of a
  raw uncompressed gel image and corresponding compressed
  images with $P=0.99, 0.975$ and $0.95$ (\textup{top}), along with corresponding posterior discovery images
(posterior probability of 1.5-fold expression, \textup{bottom}) for differences
between animals in control and long cocaine access
groups.}\label{fig:compression2}
\end{figure}

\textit{Image-based modeling using ISO-FMM.} We constructed an
isomorphic transformation for the images based on a square nonseparable
2D wavelet transform using a Daubechies wavelet with four vanishing
moments, periodic boundary conditions, and the decomposition completed
to $J=6$ frequency levels. To investigate image compression, we
generated a compression plot (Figure \ref{fig:compression_plot}) as
described in Section \ref{sec:compression}. Note that we were able to
preserve a high level of energy while retaining a small proportion of
coefficients. The top panels of Figure \ref{fig:compression2} contain
plots of one of the processed 2D gel images (uncompressed and
compressed using $P=0.99, 0.975 \mbox{ and } 0.95$) to demonstrate that
the compressed and uncompressed images look virtually identical. We
chose $P=0.975$ for our primary analyses, which modeled the $T=646
\times861 = 556{,}206$ pixels using only $T_{97.5}^*=10{,}634$ wavelet
coefficients, for a compression ratio of more than 50:1. As a
sensitivity analysis, we also ran ISO-FMM with compression levels
$P=0.95$ and $P=0.99$, yielding $T_{95}^*=4958$ and $T_{99}^*=26{,}520$
coefficients, respectively, which correspond to compression ratios of
over 100:1 and 20:1. We also considered the rectangular transform, but
found this was not as efficient in representing the 2-DE images, with
11,384 coefficients required using $P=0.975$, so we chose to use the
square transform in our analyses.

Let $Y_i(t_1, t_2), i=1, \ldots, 53$, be the $\log_2$-transformed
preprocessed gel images.
We used the following functional mixed model for these data:
\begin{equation}\label{eq:example}
Y_i(t_1,t_2) = \sum_{a=1}^3 X_{ia} B_a(t_1, t_2) + \sum_{b=1}^{21}
Z_{ib} U_b(t_1, t_2) + E_i(t_1, t_2),
\end{equation}
where $X_{ia}=1$ if gel $i$ is from an animal in group $a$, 0
otherwise, with the groups
labeled as $a=1$ control animals, $a=2$ animals with short access to
cocaine, and $a=3$ animals with long
access to cocaine. The fixed effect image $B_a(t_1, t_2)$ represents
the average gel for group $a$. The random
effects were included to model correlation between gels from the same
animal, with $Z_{ib}=1$
if gel $i$ is from animal $b$, and $U_b(t_1, t_2)$ as the random effect
image for animal $b$.
We assumed the random effect functions and residual error functions
were i.i.d. ($P=R=I$).

After transforming the images to the wavelet space, we fit the
wavelet-space version of
(\ref{eq:example}) as described in Section \ref{subsec:image_FMM}.
Maximum likelihood estimates
were used for starting values of the MCMC, and vague, proper inverse
gamma priors were assumed for the
variance components, centered on the ML estimates with information
equivalent to a sample size of $2$.
After a~burn-in of 1000, we ran the MCMC for 20,000 samples, keeping
every 10th observation. Run on
a single Xeon 2.66 GHz processor, this analysis took a total of 16.8
hours when run with $P=0.975$
compression, and 10.8 and 34.8 hours, respectively, when run with
$P=0.95$ and $P=0.99$ compressions.
Because the method is roughly linear in $T^*$, the number of basis
coefficients modeled [Herrick and Morris (\citeyear{HerMor2006})]
had we not used compression, this method would have taken approximately
600 hours using a single processor
to obtain the same number of posterior samples. We note that parallel
processing on a network or cluster
could have been used to greatly reduce the run time. Using the 2D-$\operatorname{IDWT}$,
the posterior samples of the fixed
effects in the transformed space were projected back into the data
space to yield posterior
samples of the fixed effect images $B_a(t_1, t_2)$.

Next, we constructed posterior samples for the overall mean gel image,
$M(t_1, t_2)=1/3 \{B_1(t_1, t_2) + $ $B_2(t_1, t_2) + B_3(t_1, t_2)\}$,
%(right panel of Figure \ref{fig:mean_gels}),
and to consider
contrasts corresponding to the various between-group comparisons. Here
we focus on the
image corresponding to the difference between the control group and the long
cocaine access group, $C_{13}(t_1, t_2)=B_1(t_1, t_2)-B_3(t_1, t_2)$
(upper right panel of Figure~\ref{fig:FMM_results}). Regions of
$C_{13}(t_1, t_2)$
with large negative values correspond to regions with greater protein
expression for
animals given a long access to cocaine. Regions with large positive values
correspond to regions with greater protein expression for the control animals.
Using the approach described in Section \ref{subsec:FDR}, we sought to
identify regions of the gel with at least 1.5-fold
difference between groups [$\delta=\mbox{log}_2(1.5)=0.5850$], while
controlling the
FDR at $\alpha=0.10$.

\textit{Spot-based modeling using Pinnacle.} To compare our ISO-FMM image-based
approach with a standard spot-based method, we applied \textit{Pinnacle}
[Morris, Clark and Gutstein (\citeyear{MCG2008})] to these data. First, we aligned
and preprocessed
the images, exactly as described above, to make sure that any
difference in results was not due to preprocessing but
due to the spot vs. image-based approach. Applying Pinnacle, we computed
the raw mean processed gel, and denoised it using an undecimated wavelet-based
approach. We detected spots based on their pinnacles, defined as any pixel
that is a~local maxima in both the horizontal and vertical directions
of the
wavelet-denoised average gel whose normalized intensity is greater than the
$75$th percentile on the gel. Using the Pinnacle graphical user interface,
we hand-edited the spot detection to remove obvious artifacts, and were
left with a total of 752 detected spots.
%The left panel of Figure \ref{fig:mean_gels}
%contains the raw mean gel for this data set, with detected spots
%(pinnacles)
%marked with an ``x.''
For each gel, we quantified each spot using the maximal
normalized intensity within a $5 \times5$ square around the detected pinnacle,
and then averaged intensities over replicate gels from the same animal,
yielding a $21 \times752$ matrix containing normalized spot quantifications
for each of 752 detected spots for the 21 animals. Using this matrix, we
performed $t$-tests for each pinnacle to compare the samples from animals
in the
control and long cocaine access groups, and then forwarded the
$p$-values into the \textit{fdrtool} method [Strimmer (\citeyear{Strimmer2008})] to
obtain the corresponding $q$-values, or local false
discovery rates.

%s4.2 ###
\subsection{Results} \label{subsec:example_results}

\begin{figure}

\includegraphics{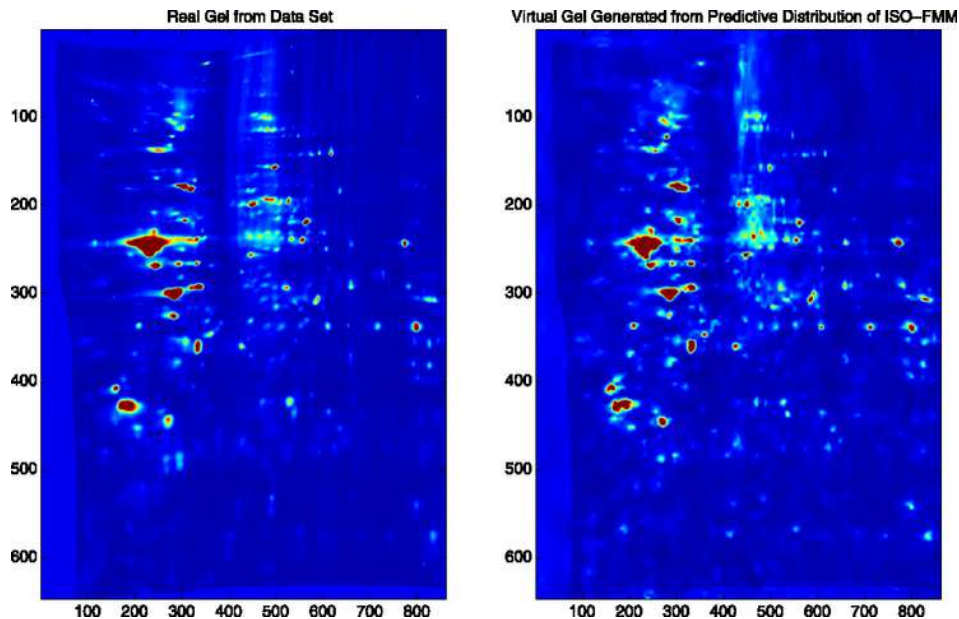}

  \caption{Virtual Gel Plot of single gel from data
example \textup{(left panel)} and a virtual gel \textup{(right panel)}, found
by sampling randomly from the posterior predictive distribution of
the ISO-FMM used to fit the sample data.  Note that the ISO-FMM is
able to sufficiently capture the structure of real 2D gels so that
the virtual gel looks very much like a real gel that could have come
from the example data set.} \label{fig:virtual_gel}
\end{figure}

\textit{Results of ISO-FMM image-based analysis.} First, to assess
whether the model
was flexible enough to model the 2-DE data, we generated a ``virtual
gel'' by
sampling from the posterior predictive distribution for the specified ISO-FMM
(\ref{eq:example}), plotted in the right panel of Figure \ref
{fig:virtual_gel} along
with an actual gel (left panel). The virtual gel looks remarkably like
a real gel,
indicating the ISO-FMM with square 2D wavelet-based modeling is able
to capture the salient features of the gel, and demonstrating the
flexibility of
this nonparametric modeling approach.

\begin{figure}

\includegraphics{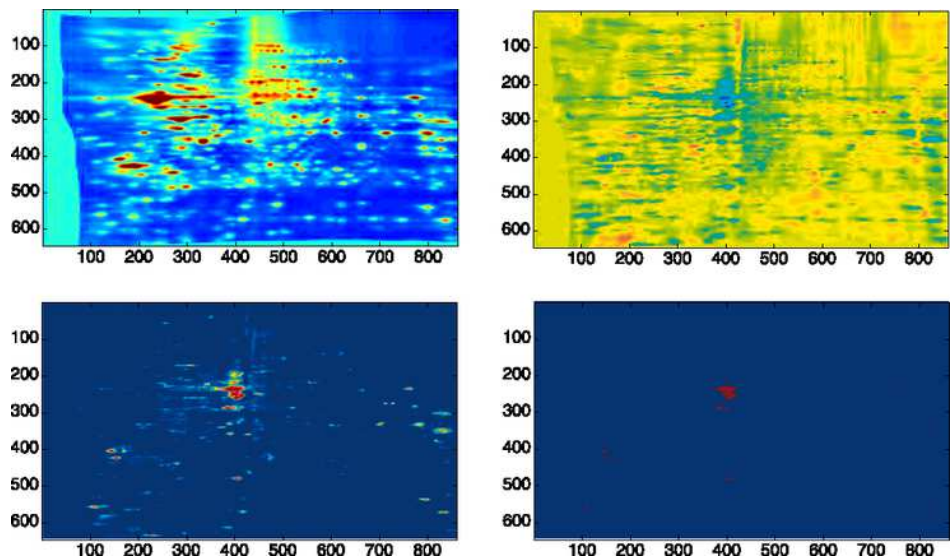}

  \caption{ISO-FMM results: Heatmaps of posterior mean of overall mean
  gel [$M(t_1,t_2)$, \textup{upper left}] and control vs. long cocaine access
  effect gel [$C_{13}(t_1,t_2)$, \textup{upper right}], plus probability discovery
plot [$p^{1.5}(t_1,t_2)$, \textup{lower left}] and regions of gel flagged as
significant (FDR${}<0.10$, 1.5-fold, \textup{lower right}).  Higher intensities are
indicated by hotter colors, lower intensities by cooler colors.}\label{fig:FMM_results}
\end{figure}

%Figure \ref{fig:mean_gels} contains the raw mean gel and the posterior
%mean
%estimate of the overall mean gel from the ISO-FMM, $M(t_1, t_2)$. Note
%that
%the model-based mean gel and raw mean gels are very similar.
Figure \ref{fig:FMM_results} summarizes the overall results of the
ISO-FMM model fitting. The top panels contain the
posterior means for the overall mean gel image $M(t_1, t_2)$ and the
control vs. long
access contrast image $C_{13}(t_1, t_2)$. In the contrast image, blue
regions correspond
to regions of the gel with higher protein expression for animals in the
long cocaine access
group; red and orange regions indicate higher protein expression for
control animals;
and yellow regions indicate no difference. Note that we see a mix of
blue and red regions, and most of
these regions resemble protein spots. This is what we would expect to
see in well-run
gel studies with differentially expressed protein spots. If most of the
effects were
all in the same direction (blue or red), or if the regions were
irregular and not spot
shaped, then we might suspect that the results were driven by some
artifacts in the data,
for example, background artifacts, which might indicate a problem in
the experiment.

The bottom left panel of Figure \ref{fig:FMM_results} is the \textit
{probability discovery plot},
$p_{13}^{1.5}(t_1,\break t_2)$, measuring the posterior probability of at
least 1.5-fold expression
differences between animals in the control and long cocaine access
groups with red regions having the highest
posterior probabilities. The bottom panels of Figure \ref
{fig:compression2} contain this
probability discovery plot for the different compression levels, and
demonstrate that
the results are robust to the choice of compression level $\cP$.
Again, these regions of high probability are
shaped like protein spots, as we would expect if they were marking
differentially expressed proteins.
Applying the FDR${}<0.10$ criterion as described in Section~\ref
{subsec:FDR}, we flagged all pixels
$(t_1, t_2)$ with $p_{13}^{1.5}(t_1, t_2)>\phi^{1.5}_{0.10}=0.757$ as
differentially expressed.
These regions are marked in red in the bottom right panel of Figure~%
\ref{fig:FMM_results}.
There are 27 contiguous regions flagged, which are summarized in Table~%
\ref{table:ISO-FMM_results}.
This image could be used to inform the spot-picker
which physical regions of the gel to cut out for MS--MS analysis to
ascertain the corresponding
protein identities. Unfortunately, for this study, the original
physical gels were no longer
available, so we could not perform the experiment to discern the
corresponding protein identities.

\begin{table}
\caption{Results of image-based ISO-FMM analysis:
Details for regions flagged as differentially expressed in
image-based ISO-FMM analysis, including coordinates of center
of region ($x_{\mathrm{FMM}},y_{\mathrm{FMM}}$), maximum posterior probability of
1.5-fold change within region ($p_{1.5}$), coordinates
of nearest detected pinnacle ($x_{\mathrm{Pinn}},y_{\mathrm{Pinn}}$) and corresponding
p-value (pval) and fold-change (FC)}\label{table:ISO-FMM_results}
\tabcolsep=0pt
\vspace*{-3pt}
\begin{tabular*}{\textwidth} {@{\extracolsep{\fill}}lcd{2.4}ccd{2.4}cc@{}}
\hline
$\bolds{x_{\mathrm{FMM}}}$ & $\bolds{y_{\mathrm{FMM}}}$ & \multicolumn{1}{c}{$\bolds{p_{1.5}}$}
& $\bolds{x_{\mathrm{\mathrm{Pinn}}}}$ & $\bolds{y_{\mathrm{Pinn}}}$ & \multicolumn{1}{c}{\textbf{pval}} & \textbf{FC} & \textbf{Comments} \\
\hline
403&263&{>}0.9995&406&264&0.0004&2.6932&Found by both methods\\
415&257&{>}0.9995&418&257&0.0003&2.1518&Found by both methods\\
410&239&{>}0.9995&410&239&0.0220&1.8648&Found by both methods\\
393&239&0.9995&393&239&0.0008&2.1049&Found by both methods\\
401&252&0.9995&405&252&0.0001&1.7319&Found by both methods\\
386&290&0.9890&381&291&0.0062&2.2087&Found by both methods\\
405&483&0.9785&407&483&0.0017&1.7544&Found by both methods\\
398&291&0.9150&407&296&0.0015&1.5016&Found by both methods\\
405&203&0.8660&405&203&0.0025&1.8578&Found by both methods\\
391&360&0.8240&389&360&0.0009&1.8170&Found by both methods\\
343&227&0.8210&341&228&0.0796&1.8084&Found by both methods\\
712&279&0.8035&711&282&0.0168&1.5134&Found by both methods\\
727&278&0.7590&728&280&0.0103&1.9038&Found by both methods\\
831&557&0.8925&835&575&0.0013&1.4495&Fold change too small\\
399&220&0.8865&402&222&0.0036&1.4560&Fold change too small\\
109&560&0.8630&120&559&0.0167&1.2412&Fold change too small\\
232&639&0.8560&240&639&0.0087&1.4723&Fold change too small\\
797&540&0.8210&799&541&{<}0.0001&1.4866&Fold change too small\\
388&335&0.7820&383&333&{<}0.0001&1.4096&Fold change too small\\
832&351&0.8540&828&357&0.1466&1.2377&In right tail of major spot\\
762&238&0.8375&773&243&0.7220&1.0500&In left tail of major spot\\
144&408&0.9815&160&409&0.9883&1.0011&In left tail of major spot\\
154&427&0.9775&177&427&0.3312&1.0477&In left tail of major spot\\
704&332&0.8795&713&338&0.4748&1.0767&In left tail of major spot\\
559&222&0.9420&562&220&0.4697&1.0558&Between two visible spots\\
449&251&0.8940&446&257&0.1105&1.2458&Between two visible spots\\
308&175&0.7985&312&172&0.0471&1.3055&Between two visible spots\\
\hline
\end{tabular*}
\vspace*{-6pt}
\end{table}

\textit{Results of Pinnacle spot-based analysis.}
Performing a spot-based analysis, we flagged a spot as significant if
its $q$-value was less than 0.10, and
the effect size was at least $\log_2(1.5)=0.5850$, indicating at least
1.5-fold difference. This led to 17 differentially expressed spots
between animals in the control and long cocaine access groups,
%marked with a ``o'' in Figure \ref{fig:mean_gels}, and
which are summarized in Table \ref{table:Pinn_results}.

\begin{table}
\caption{Results of spot-based Pinnacle analysis: Details for spots flagged
as differentially expressed in spot-based Pinnacle analysis,
including location ($x$, $y$), $p$-value (pval), $q$-value (qval) and fold-change (FC).
Also included is the maximum $p_{1.5}(t_1,t_2)$ from the ISO-FMM
within a 5-by-5 neighborhood around the corresponding Pinnacle}\label{table:Pinn_results}
\vspace*{-3pt}
\begin{tabular*}{\textwidth} {@{\extracolsep{\fill}}ld{3.0}d{2.3}ccd{2.3}c@{}}
\hline
$\bolds{x}$ & \multicolumn{1}{c}{$\bolds{y}$} & \multicolumn{1}{c}{\textbf{pval}} &\textbf{qval}
& \textbf{FC} & \multicolumn{1}{c}{$\bolds{p_{1.5}}$} & \textbf{Comments} \\
\hline
410&239&0.002&0.008&1.865&{>}0.999&Found by both methods\\
418&257&{<}0.001&0.002&2.152&{>}0.999&Found by both methods\\
406&264&{<}0.001&0.003&2.693&{>}0.999&Found by both methods\\
405&252&{<}0.001&0.001&1.732&0.999&Found by both methods\\
393&239&0.001&0.004&2.105&0.999&Found by both methods\\
381&291&0.006&0.014&2.209&0.989&Found by both methods\\
407&483&0.002&0.007&1.754&0.979&Found by both methods\\
407&203&0.005&0.013&1.671&0.866&Found by both methods\\
389&360&0.001&0.005&1.817&0.824&Found by both methods\\
341&228&0.080&0.068&1.808&0.821&Found by both methods\\
711&282&0.017&0.027&1.513&0.804&Found by both methods\\
407&296&0.001&0.007&1.502&0.788&Found by both methods\\
728&281&0.014&0.024&1.638&0.759&Found by both methods\\
379&263&0.009&0.018&1.595&0.487&Just missed threshold\\
257&60&0.062&0.048&1.663&0.463&Just missed threshold\\
409&163&0.006&0.014&1.504&0.160&FC barely above 1.5\\
798&177&0.004&0.012&1.543&0.019&FC barely above 1.5, faint spot\\
\hline
\end{tabular*}
\vspace*{-7pt}
\end{table}

\textit{Comparison of Pinnacle and ISO-FMM results.} Note that the
Pinnacle and ISO-FMM results
are not entirely comparable since they use different criteria; Pinnacle
flags a spot as significant
if the $q$-value is less than 0.10 (based on a $t$-test with point mass
null hypothesis) and the effect
size is at least 1.5-fold, while the ISO-FMM flags a region as
significant if its posterior probability of
a 1.5-fold difference is large enough to cross the estimated $\mathit{FDR}<0.10$
significance threshold. However,
we still found it enlightening to qualitatively compare these results.

Out of the 17 spots flagged as significant by Pinnacle, 13 were
contained within regions flagged by the ISO-FMM
analysis. Two of the others have high probabilities of 1.5-fold
difference (${\approx}0.50$), but that just
missed the $\mathit{FDR}<0.10$ threshold of $\phi^{1.5}_{10}=0.757$. The other
two were very faint spots with fold
changes marginally greater than 1.5-fold.

\begin{figure}

\includegraphics{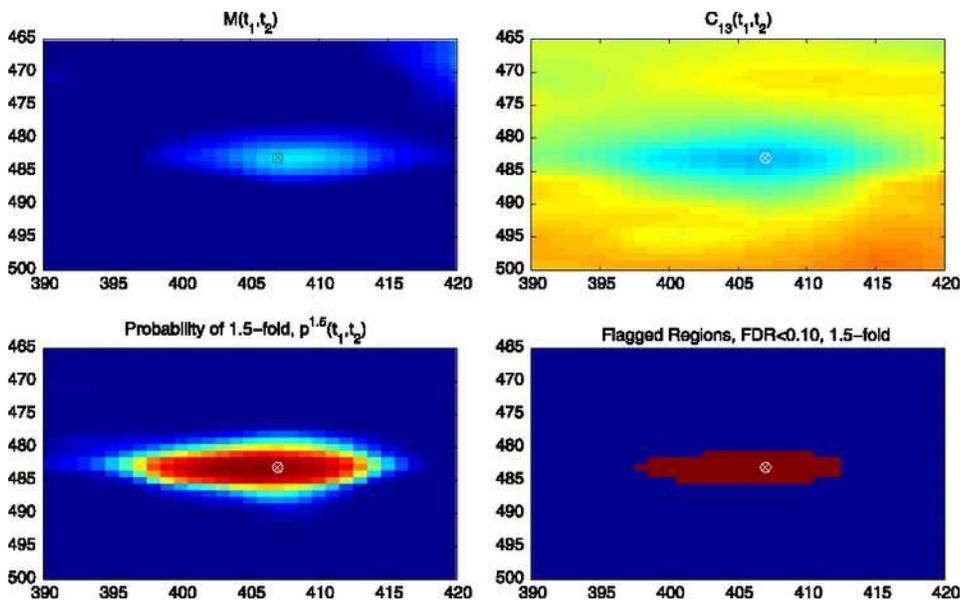}%
\vspace*{-3pt}
  \caption{Specific Results 1: Posterior mean of overall mean gel (\textup{upper left}),
  effect gel (\textup{upper right}), probability discovery plot (\textup{lower left}), and
  indicating ISO-FMM flagged regions (\textup{lower right}) for
region marked by small box in Figure \protect\ref{fig:FMM_results}, with pinnacles
for detected spots marked ($\times$), and differential expression in Pinnacle
analysis indicated by a (\textup{o}).  Note that region flagged by ISO-FMM corresponds to visible
spot also detected by Pinnacle analysis.}\label{fig:specific_results_1}
\vspace*{-7pt}
\end{figure}

Of the 27 regions flagged in the ISO-FMM analysis, 13 have
corresponding pinnacle results. Figure~\ref{fig:specific_results_1} displays one such region, marked by the
small box in Figure~\ref{fig:FMM_results}.
We see that this region precisely corresponds to the boundaries of a~single visible spot, and the Pinnacle
location is marked by the ``$\times$,'' with the ``o'' indicating it was flagged
as significant by the Pinnacle
spot-based analysis. Six more regions clearly correspond to visible
spots in the mean gel that have small $p$-values,
but estimated fold-changes less than 1.5. The remaining 8 nonmatched
regions corresponded to
subsets of detected spots or in the areas between two spots; four
corresponded to regions in the left tail of visible spot,
one to a region in the right tail of a visible spot, and the remaining
three appeared
between 2 visible spots. These results were not found by the spot-based
Pinnacle approach.

\begin{figure}

\includegraphics{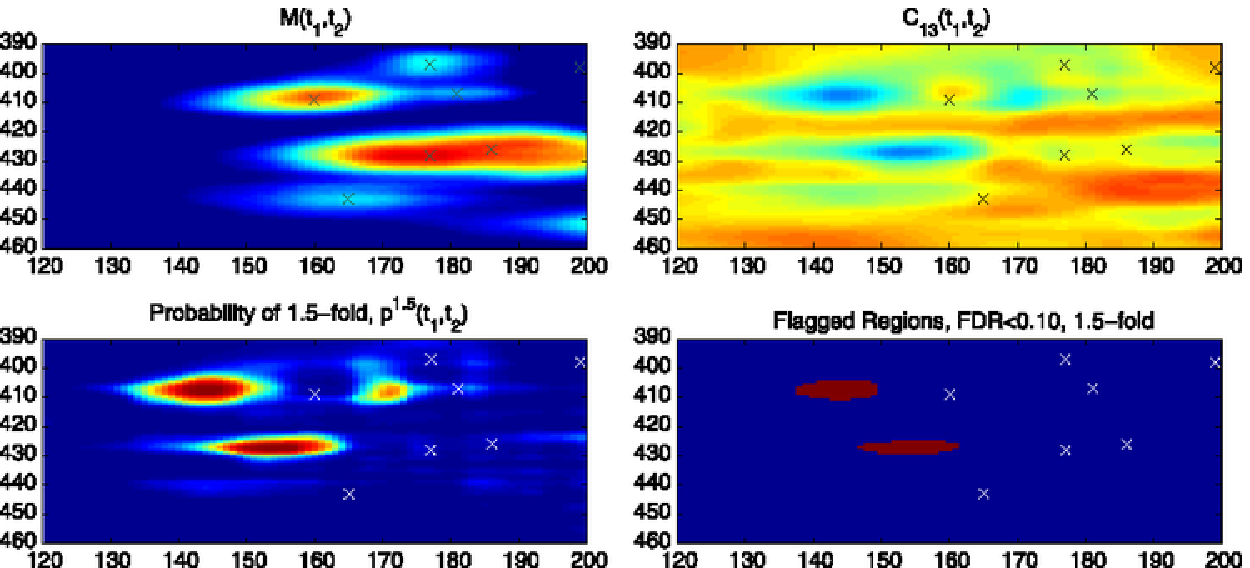}%
\vspace*{-3pt}
  \caption{Specific Results 2: Posterior mean of overall
  mean gel (\textup{upper left}), effect gel (\textup{upper right}), probability
  discovery plot (\textup{lower left}), and indicating ISO-FMM flagged
  regions (\textup{lower right}) for
region marked by large box in Figure \protect\ref{fig:FMM_results}, with
pinnacles for detected spots marked ($\times$), and differential expression
in Pinnacle analysis indicated by a (\textup{o}).  Note that regions flagged
by ISO-FMM correspond to tails
of visible spots that themselves are not differentially expressed.
These results are not found by the Pinnacle analysis.}\label{fig:specific_results_2}
\vspace*{-7pt}
\end{figure}

One interesting part of the gel containing two such regions is
indicated by the large box in Figure
\ref{fig:FMM_results}, and presented in detail in Figure \ref
{fig:specific_results_2}. From the mean gel
image, we see that this field contains 7 visible protein spots detected
by Pinnacle, as marked by the $\times$'s. From
the other panels, we see two regions flagged as differentially
expressed (long access${}>{}$control) by ISO-FMM,
and a third region with high probability of differential expression but
not quite exceeding the $\mathit{FDR}=0.10$ threshold
$\phi^{1.5}_{0.10}=0.757$. These flagged regions resemble protein
spots but do not correspond to the visible
protein spots in the mean gel. Rather, they correspond to the left
tails of the two dominant spots in this field,
which both appear to have slightly extended long tails. The spot-based
approach found no significant spots in this
region of the gel, so these discoveries would have been missed had we
not conducted the image-based analysis. Other
spots have similar behavior, and their details can be seen in
supplementary Figures 3--10 [Morris (\citeyear{Morris2010})]. A key question is what
these flagged regions could represent.

%Some of these regions corresponded to protein spots found to be
%differentially expressed by Pinnacle.
%For example, Figure \ref{fig:specific_results_1} contains the mean
%gel, contrast gel, probability
%discovery gel, and flagged region gel images for the zoomed-in portion
%of the gel indicated by the
%small white square in Figure \ref{fig:FMM_results}. This field
%contains a single spot that is marked
%as differentially expressed, with higher expression shown in the
%animals in the long cocaine access
%group than in those in the control group. The white ``x'' and ``o''
%indicate that this was
%detected as a spot by Pinnacle, and was found to be differentially
%expressed in the spot-level analysis, as well.

%The zoomed-in plots from another part of the gel, indicated by the
%large box in Figure \ref{fig:FMM_results},
%are presented in Figure \ref{fig:specific_results_2}. This part of the
%gel is very interesting.

\textit{Interpretation of results.} As mentioned in Section \ref
{sec:app}, a well-known issue in 2-DE is the presence of \textit
{co-migrating proteins},
that is, distinct proteins that visually appear to be part of the same
protein spot. Studies have shown that some spots
on a gel can have as many as 5 or 6 distinct proteins [Gygi et al.
(\citeyear{Gygietal200})]. These co-migrating proteins can be
different proteins or post-translational modifications of the same
protein, which can also be functionally
distinct. If two proteins have similar combinations of pH and molecular
mass, it is possible that the proteins will
run together in the same visible spot on the 2-DE. It can be very
difficult, and sometimes impossible, for any spot
detection method to deconvolve these multiple spots into separate
protein spots, especially if one of them has
considerably higher abundance than the others. The inability to resolve
these co-migrating proteins is one of the
key criticisms levied against 2-DE. The significant regions flagged by
ISO-FMM but not by Pinnacle which appear in the
tail of a visible spot or between two visible spots may indicate
differentially expressed co-migrating proteins visibly
masked by a more abundant non differentially-expressed protein, and that
these proteins were completely missed by the
spot-based analysis. Studies are underway to confirm this possibility.

%We are in the process of developing a simulation engine to generate
%virtual 2-DE data to use for simulation studies that models spots as
%2D Gaussians with volumes corresponding to the protein %abundances. We
%simulated a set of random gels, using the estimated distributions of
%spot locations and intensities from a real gel set as the basis for
%the simulation. Figure %\ref{fig:simulated_gel} contains one region of
%a simulated gel, with the centers of true protein spots indicated by
%``o'', and the spots detected by Pinnacle as ``x''. Note that
%co-migrating proteins %in the virtual gels tend to yield irregular
%spots or spots with long tails, and that the spot detection method
%frequently cannot resolve these co-migrating spots. These spot shapes
%are much like the %spots in Figure \ref{fig:specific_results_2} with
%long tails, and lend credence to the notion that the

In this way, the image-based modeling approach may be able to extract
more protein information from the gels
than spot-based approaches. Because spot-based modeling approaches have
been almost universally used to date,
this means that perhaps 2-DE contains more proteomic information than
was previously known. The image-based
approach may effectively increase the realized resolution of the gels
and better extract the proteomic information
they contain. Further biological studies are needed to validate these
conjectures.

%s5 ###
\section{Discussion} \label{sec:discussion}
In this paper we have discussed a general Bayesian method for
quantitative image
data based on functional mixed models that uses an isomorphic modeling
approach. The underlying FMM
framework is very general, can simultaneously model any number of
covariates, each having
their own fixed effect image of general form, and can account for
correlation between the images
using random effect images and between-image covariance matrices. The
results from the method
can be used to perform FDR-based Bayesian inference that takes both
practical and statistical
significance into account, and flags significant regions of the fixed
effect images.

Previous work on functional mixed models has been limited to single-dimen\-sional functions;
here we have shown how this approach can be applied to images of
dimension 2 or higher. Also, previous work on
functional mixed models has been based on specific modeling strategies
using smoothing splines
[Guo (\citeyear{Guo2002})] or wavelets [Morris and Carroll (\citeyear{MorCar2006})]. In this paper we
have described a
general modeling strategy that involves using an isomorphic
transformation to map the
data to an alternative space, where modeling can be done more
parsimoniously and smoothing
or regularization naturally done, and results can be mapped back to the
original data space for
final inference. This method, ISO-FMM, contains WFMM as a special case,
but can be also
applied much more generally using other isomorphic transformations.
With each proposed
transformation, careful thought needs to be given to the modeling
choices and their
implications in the data space, and thus further work is required to
apply this approach
using certain other transformations. This general modeling strategy can
be used for the
1D FMM, or for the higher dimensional FMM that is the primary interest
of this paper.
This isomorphic modeling approach is a general strategy with potential
for application
to a variety of other contexts.

We introduced a compression method to reduce the dimensionality of the
data that is
appropriate when the chosen isomorphic transformation leads to a sparse
representation
for all of the images, as is true for wavelets. This compression method
can also be
applied to any functional data, 1-D or higher. We have found it is possible
to use compression to speed up the computations by one or two orders of
magnitude, which
greatly reduces the memory overhead requirements without substantively
changing the
results.

While the method is complex, we have developed general freely
avai\-lable~software
(\href{http://biostatistics.mdanderson.org/SoftwareDownload/SingleSoftware.aspx?Software\_Id=70}%
{http://biostatistics.mdanderson.org/SoftwareDownload/}\break
\href{http://biostatistics.mdanderson.org/SoftwareDownload/SingleSoftware.aspx?Software\_Id=70}%
{SingleSoftware.aspx?Software\_Id=70})
to implement it that is efficient enough to handle very large data sets
and complex models,
and is relatively straightforward to use, considering the complexity of
the method. If
the user is satisfied with automatic vague proper priors and default
wavelet bases, then
the method can run completely automatically if the user simply
specifies the~$Y$, $X$ and
$Z$ matrices. Users who wish to use alternative basis functions can
compute the
$D$ matrix themselves and feed that into the program, indicating which
groups of coefficients
will share common smoothing parameters. The code automatically
generates posterior means and
quantiles (default 0.005, 0.01, 0.025, 0.975, 0.99, 0.995) for prespecified
contrasts involving the
$B_a(t_1, t_2)$, plus the posterior probabilities of specific effect
sizes $p^{\delta}_a(t_1, t_2)$
for specified choices of $\delta$ \{default $\delta=\log_2(1.25),
\log_2(1.5), \log_2(2.0)$\}.
Thus, plots like those generated in Figure \ref{fig:FMM_results} can
be quickly generated once
the method is run. The code is continually being updated as the scope
of the FMM framework is
extended, so more features will be added in the future, as will an R
interface for running the
method.

Applying this method to 2-DE data, we found that the ISO-FMM was able
to find differentially expressed proteins that may be co-migrating
proteins that
would not have been found using the usual spot-based analysis
approaches. The ISO-FMM may be capable of extracting
more proteomic information from the gels than was previously known to
be there.
This method can easily be applied to DIGE data by just modeling
$Y_i(t_1, t_2)$
to be the log ratio of the two channels. Although this paper focused on 2-DE,
the ISO-FMM can be applied to any type of quantitative image data,
including fMRI, LC--MS
and other commonly encountered applications. This rigorous, automated
method can
be useful in extracting information and performing inference
for quantitative image data.

\section*{Acknowledgments}
We thank
Andrew Dowsey and Brittan Clark for
aligning the gels, and George Koob from Scripps Research Institute
for helpful advice. We also thank Lee Ann Chastain for excellent
editorial assistance.

\begin{supplement}[id=suppA]
\sname{Supplement A}
\stitle{Computational details for wavelet-space implementation of ISO-FMM
for image data}
\slink[doi]{10.1214/10-AOAS407SUPPA}
\slink[url]{http://lib.stat.cmu.edu/aoas/407/supplement\_A.pdf}
\sdescription{Computational details for wavelet implementation of the
ISO-FMM for image data, including empirical Bayes method for estimating
regularization parameters, MCMC details and Metropolis--Hastings
details for covariance parameters.}
\end{supplement}

\begin{supplement}[id=suppB]
\sname{Supplement B}
\stitle{Supplementary figures}
\slink[doi]{10.1214/10-AOAS407SUPPB}
\slink[url]{http://lib.stat.cmu.edu/aoas/407/supplement\_B.pdf}
\sdescription{Supplementary figures, including a virtual 2d gel
simulated from the model, a~demonstration of the spatial covariance
structure induced by the model and 8 plots containing zoomed-in results
from analysis of application data in certain interesting regions of the gel.}
\end{supplement}

\begin{supplement}[id=suppC]
\sname{Supplement C}
\stitle{Spatial covariance structure in image WFMM\\}
\slink[doi]{10.1214/10-AOAS407SUPPC}
\slink[url]{http://lib.stat.cmu.edu/aoas/407/supplement\_C.pdf}
\sdescription{Basic illustration of spatial covariance structure
induced by ISO-FMM with 2D wavelet transforms and independence assumed
in the wavelet space. Basic demonstration described, and some plots
provided. Movie file spatial\_covariance.wvm also available as
supplementary material to further illustrate these results.}
\end{supplement}

\begin{supplement}[id=suppD]
\sname{Supplement D}
\stitle{Movie file illustrating spatial covariance structure of
ISO-WFMM with 2D wavelet transform}
\slink[doi]{10.1214/10-AOAS407SUPPD}
\slink[url]{http://lib.stat.cmu.edu/aoas/407/supplement\_D.wmv}
\sdescription{Windows movie file illustrating the nonstationary
spatial covariance structure induced by the ISO-FMM with 2D wavelet
bases, with independence assumed among wavelet coefficients.
Description of data yielding this movie is provided in the file
``Spatial Covariance Structure in Image WFMM.pdf,'' also available as
supplementary material.}
\end{supplement}

%suskaldyti doi

\printaddresses

\end{document}